\documentstyle[12pt]{article}
\large
\begin{document}
\parindent=0.8cm
\vspace{1cm}

\begin{center}
{\LARGE \bf Macroscopic Quantum Coherence in Small\\ 
Antiferromagnetic Particle and the Quantum Interference  Effects}
\end{center}
\begin{center}{ \large Yi-Hang Nie$^{a,c}$ , Y.-B Zhang$^{a,b}$ , J.-Q Liang$^{a,b}$ ,\\ 
H.J.W.M$\ddot{{\rm u}}$ller-Kirsten$^{d}$ and F.-C Pu$^{b,e }$}\\
\end{center}
\begin{center}
 $^{a}${\small Institute of Theoretical Physics, Shanxi University , Taiyuan, 
Shanxi 030006, China }\\
$^{b}${\small Institute of Physics and Center for Condensed Matter Physics, Chinese 
Academy of Sciences,\\ Beijing 100080, China \\}
$^{c}${\small Department of Physics , Yanbei Normal Institute, Datong, Shanxi 
037000, China}\\
$^{d}${\small Department of Physics, University of Kaiserslautern, 67653 
Kaiserslautern, Germany}\\
$^{e}${\small Department of Physics, Guangzhou Normal College, Guangzhou 510400, 
China}\\
 \end{center}

\begin{abstract}
Starting from the Hamiltonian operator of the  noncompensated two-sublattice 
model of a small antiferromagnetic particle, we derive the effective Lagrangian
of a biaxial antiferromagnetic particle in an external magnetic field with the help 
of spin-coherent-state path integrals. Two unequal level-shifts induced by 
tunneling through  two types of barriers are obtained using the  instanton 
method. The energy spectrum is found from Bloch theory regarding the periodic 
potential as a superlattice. The external magnetic field indeed removes 
Kramers' degeneracy, however a new quenching of the energy splitting depending 
on the applied magnetic field is observed for both integer and half-integer
 spins due to the quantum interference between transitions through two types 
 of barriers.
\begin{flushleft}
PACS number(s): 75.10.Jm, 75.30.Ee, 03.65.Sq  

\end{flushleft}
\end{abstract}

\section
{Introduction}

The magnetization vector in solids is traditionally viewed as a classical variable. 
In recent years, theoretical and experimental works have demonstrated, however, 
that the vector can tunnel quantum mechanically out of  metastable magnetic states or  
resonate between two degenerate ground states[1-13] known as macroscopic 
quantum phenomena(MQP) which are distinguished  
into the  macroscopic quantum tunneling (MQT) and the macroscopic quantum  
coherence (MQC) respectively. Quantum tunneling of the magnetization 
vector in small single-domain ferromagnetic (FM) particles[1,2],  quantum 
nucleation of FM bubbles[3], and quantum depinning of a domain wall
in bulk ferromagnets[4] are typical examples of the macroscopic quantum
phenomena. Similar effect also exists in small 
single-domain antiferromagnetic (AFM) particles in which the N$\acute{\rm e}$el 
vector plays a role of macroscopic variable and  can tunnel between  
orientations of lowest energy[5,6]. Since the 
tunneling rate in the AFM particles is much higher than that in the FM particles  
[7], the AFM particles are expected to be  a better candidate for the 
observation  of MQP than the FM particles. Another interesting  phenomenon 
relating to tunneling in magnetization is that for spin systems with discrete rotation symmetry
of two folds, the tunneling rate is completely  suppressed for half-integer total 
spin known as Kramers' degeneracy[8]. Such an effect is called topological quenching in literature
[9] and has been  studied extensively[8-14].

In literature the AFM particle is usually described by the  N$\acute{\rm e}$el 
vector of two collinear sublattices whose magnetizations are coupled by strong  
exchange interaction. External magnetic field does not play a role since the 
net magnetic moment vanishes for idealized sublattices. The quantum and classical
transitions of the N$\acute{\rm e}$el vector in antiferromagnets has been well
studied[15] in terms of the idealized sublattice model. The temperature
dependence of quantum tunneling was also given for the same model[16] and the
theoretical result agrees with the experimental observation[17]. A biaxial
 AFM particle
with a small non-compensation of sublattices in the absence of external magnetic 
field was studied in Ref.[18] where it was shown that the noncompensated magnetic 
moment leads to a modification of oscillation frequency around the equilibrium 
orientations of the  N$\acute{\rm e}$el vector. In the present paper we 
demonstrate that the small noncompensated magnetic moment  obtains  extra 
energy in magnetic field which  changes the original equilibrium orientations 
of the  N$\acute{\rm e}$el vector and results in interesting  tunneling effects 
in AFM particles. With the help of spin-coherent-state  
path integrals we convert the spin system into a pseudoparticle moving in a 
effective potential $-V(\phi)$ with a periodically recurring asymmetric twin 
barriers which lead to two kinds of instantons. The total effect of tunneling
gives rise to the level splitting which is determined with Bloch theory regarding 
the periodic potential as a superlattice. We show explicitly that the external 
magnetic field removes the rotation  symmetry of two-fold and , therefore, 
the Kramers' degeneracy[10]. The level splitting is not quenched any longer
for half-integer spin. However a remarkable observation is that quantum 
interference between transitions through  two-type of  potential barriers 
results in an oscillation of level splitting with the external field. The 
splitting  could be entirely suppressed at the certain value of magnetic field 
 due to the disconstructive interference.

\section
{The effective Lagrangian of a biaxial AFM particle  with a small 
non-compensated magnetic moment and the equilibrium orientations of the  
N$\acute{\rm e}$el vector }

Consider a biaxial AFM particle having two collinear FM sublattices with a  
small non-compensation.We assume that the particle possesses 
a ${\bf x}$ easy axis and  x-y easy plane, and the magnetic field  ${H}$
is applied along the ${\bf y}$ direction. Regarding each sublattice as a FM 
particle the Hamiltonian operator  of the AFM particle has the form
\begin{equation}
\hat {\rm H}=\sum_{\alpha=1,2}(k_{\bot}{\hat{S}}^{z2}_{\alpha}+k_{\|}{ \hat
{S}}^{y2}_{\alpha}-\gamma HS^{y}_{\alpha})+J\hat{ S}_{1}\cdot
\hat{S}_{2},
\end{equation}
where $k_{\bot}$, $k_{\|}> 0$ are the anisotropy constants, $J$ is the exchange 
constant, $\gamma$ is the gyromagnetic ratio, and $\hat {S}_{1}\; , \;  
\hat{S}_{2}$ denote the spin operators in two sublattices with the commutation  relation 
$[{S}^{i}_{\alpha}, {S}^{j}_{\beta}]=i\hbar \epsilon_{ijk} \delta_  
{\alpha \beta}\hat{S}^{k}_{\alpha}\;(i,j,k=x,y,z\;;\;\alpha, \beta=1,2)$. 
Here we emphasize that because of the non-compensation of the two collinear FM 
sublattices the interaction terms with magnetic field, $i.e.$ the third term in the 
summation  of Eq.(1), do not vanish and result in the equilibrium-orientation 
change of the  N$\acute{\rm e}$el vector. We begin  
with the evaluation of the matrix element of the evolution operator in spin- 
coherent-state representation by means of the coherent state path integrals 
\begin{equation}
\langle{\bf N}_{f}|{\rm e}^{-2i\hat{\rm H}T/\hbar}|{\bf N}_{i}\rangle=\int \left[ 
\prod^{M-1}_{k=M}{\rm d}\mu({\bf N}_{k}) \right]  \left[ \prod^{M}_{k=1}\langle 
 {\bf N}_{k}|{\rm e}^{-i\epsilon \hat{\rm H}/\hbar}|{\bf N}_{k-1}\rangle \right].
\end{equation}
Here we define $| {\bf N}\rangle=|{\bf n}_{1} \rangle |{\bf n}_{2}\rangle\;,\; 
|{\bf N}_{M}\rangle=|{\bf N}_{f}\rangle=|{\bf n}_{1,f}\rangle|{\bf n}_{2,f}
\rangle\;,\;|{\bf N}_{0}\rangle=|{\bf N}_{i}\rangle=|{\bf n}_{1,i}\rangle
|{\bf n}_{2,i} \rangle$ and $t_{f}-t_{i}=2T\;,\; \epsilon=2T/M$,respectively. 
The spin coherent state is defined as
\begin{equation}
|{\bf n}_{\alpha}\rangle={\rm e}^{-i\theta_{\alpha}\hat{A}_{\alpha}}|S_{
\alpha},S_{\alpha}\rangle\;,\;\;\;(\alpha=1,2),
\end{equation}
where ${\bf n}_{\alpha}=(\sin \theta_{\alpha}\cos \phi_{\alpha}, \sin
\theta_{\alpha}\sin \phi_{\alpha},\cos \theta_{\alpha})$ is the unit  
vector, $\hat {A}_{\alpha}=\sin \phi_{\alpha} \hat{S}^{x}_{\alpha}-\cos 
\phi_{\alpha} \hat{S}^{y}_{\alpha}$ and $|S_{\alpha},S_{\alpha}\rangle$ is
the reference spin eigenstate. The measure is defined by
\begin{equation}
{\rm d}\mu({\bf N}_{k})=\prod_{\alpha=1,2}\frac{2S_{\alpha}+1}{4\pi}{\rm d}
{\bf n}_{\alpha,k}\;,\;\;\;\;{\rm d}{\bf n}_{\alpha,k}=\sin \theta_{\alpha,k}
{\rm d}\theta_{\alpha,k}{\rm d}{\phi}_{\alpha,k} .
\end{equation}
As $M\rightarrow \infty\;,\;\epsilon \rightarrow 0$, $\exp (-i\epsilon
\hat{\rm H}/\hbar) \approx 1- i\epsilon \hat{\rm H}/\hbar$, with
\begin{eqnarray}
\hat{\rm H} &=& \sum_{\alpha=1,2} \left[ \frac{k_{\|}}{2} \hat{S}^{2}_{\alpha}+
\left(k_{\bot}-\frac{k_{\|}}{2} \right) \hat{S}^{z2}_{\alpha}-\frac{k_{\|}}{4}
\left( \hat{S}^{+2}_{\alpha}+\hat{S}^{-2}_{\alpha} \right)-\frac{\gamma H}{2i} 
\left( \hat{S}^{+}_{\alpha}-\hat{S}^{-}_{\alpha} \right) \right]  \nonumber \\
	& &+ \frac{J}{2} \left[ \left( \hat{S}^{-}_{1} \hat{S}^{+}_{2}+
	      \hat{S}^{+}_{1} \hat{S}^{-}_{2}\right)+2\hat{S}^{z}_{1} 
	      \hat{S}^{z}_{2} \right],
\end{eqnarray}
where $\hat{S}^{+}_{\alpha}=\hat{S}^{x}_{\alpha}+i\hat{S}^{y}_{\alpha}\;,\; 
\hat{S}^{-}_{\alpha}=\hat{S}^{x}_{\alpha}-i\hat{S}^{y}_{\alpha}$. Making
use of the following approximation
\begin{equation}
\langle{\bf N}_{k}|\hat{\rm H}|{\bf N}_{k-1}\rangle\approx \langle {\bf N}_{k}|
\hat{\rm H}|{\bf N}_{k}\rangle  \langle {\bf N}_{k}| {\bf N}_{k-1}\rangle
\end{equation}
with
\begin{eqnarray}
\langle {\bf N}_{k}|{\bf N}_{k-1} \rangle &=& \prod_{\alpha=1,2} \langle{\bf n}
_{\alpha,k}|{\bf n}_{\alpha,k-1} \rangle,    \\
\langle {\bf N}_{k}|\hat{\rm H}|{\bf N}_{k} \rangle &=& \sum_{\alpha=1,2}[S^{2}
  _{\alpha}(k_{\bot} \cos^{2} \theta_{\alpha,k}+k_{\|} \sin^{2} \theta
  _{\alpha,k} \sin^{2} \phi_{\alpha,k}) -\gamma HS_{\alpha} \sin 
  \theta_{\alpha,k} \sin \phi_{\alpha,k}]  \nonumber\\
      & &+ JS_{1}S_{2}[\sin \theta_{1,k} \sin \theta_{2,k} \cos (\phi
      _{1,k}- \phi_{2,k})+\cos \theta_{1,k} \cos \theta_{2,k}],  \\
\langle {\bf n}_{\alpha,k}|{\bf n}_{\alpha,k-1} \rangle &=& \left( \frac{1+
      {\bf n}_{\alpha,k} \cdot {\bf n}_{\alpha,k-1}}{2} \right)^{S_{\alpha}}
      \exp [-iS_{\alpha}A_{\alpha}({\bf n}_{\alpha,k}, {\bf n}_{\alpha,k-1}, 
      {\bf n}_{0})] \nonumber   \\
      & \approx & \exp[-iS_{\alpha}(\phi_{\alpha,k}-\phi
      _{\alpha,k-1})(1-\cos \theta_{\alpha,k})],
\end{eqnarray}
and $A_{\alpha}({\bf n}_{\alpha,k}, {\bf n}_{\alpha,k-1}, {\bf n}_{0})$ being 
the area of the spherical triangle with vertices[19] at ${\bf n}_{\alpha,k}\;,\;
{\bf n}_{\alpha,k-1}$ and ${\bf n}_{0}=(0,0,1)$. Under the large $S$ 
limit we obtain
\begin{equation}
\langle{\bf N}_{f}|{\rm e}^{-i\epsilon \hat{\rm H}T/ \hbar}|{\bf N}_{i}\rangle=
{\rm e}^{-i\sum_{\alpha=1,2}S_{\alpha}(\phi_{\alpha,f}-\phi_{\alpha,i})}
\int\prod_{\alpha=1,2}{\cal D}[\phi_{\alpha}] {\cal D}[\theta_{\alpha}]
\exp \left( \frac{i}{\hbar}\int^{t_{f}}_{t_{i}} {\cal L} {\rm d}t  \right).
\end{equation}
Where ${\cal L}={\cal L}_{0}+{\cal L}_{1}$ is the Lagrangian with
\begin{eqnarray}
{\cal L}_{0}&=&\sum_{\alpha=1,2}S_{\alpha} \dot{\phi}_{\alpha} \cos 
\theta_{\alpha}-JS_{1}S_{2}[\sin \theta_{1} \sin \theta_{2} \cos
(\phi_{2}-\phi_{1})+\cos \theta_{1} \cos \theta_{2}],  \\
{\cal L}_{1}&=&-\sum_{\alpha=1,2}(k_{\bot}S^{2}_{\alpha} \cos^{2} \theta
_{\alpha}+ k_{\|}S^{2}_{\alpha} \sin^{2} \theta_{\alpha} \sin^{2} \phi
_{\alpha} + \gamma HS_{\alpha} \sin \theta_{\alpha} \sin \phi
_{\alpha}).
\end{eqnarray}
For our interest of quantum transition between macroscopic states only 
the low energy trajectories with almost antiparallel $S_{1}$ and $S_{2}$
contribute to the path integral[18]. We therefore replace $\theta_{2}$ 
and $\phi_{2}$ by $\theta_{2}=\pi-\theta_{1}-\epsilon_{\theta}$ and 
$\phi_{2}=\pi+\phi_{1}+\epsilon_{\phi}$, where $\epsilon_{\theta}$ and 
$\epsilon_{\phi}$ denote small fluctuations. Working out the fluctuation 
integrations over  $\epsilon_{\theta}$ and $\epsilon_{\phi}$ the transition 
amplitude Eq.(10) reduces to
\begin{eqnarray}
\langle {\bf N}_{f}|{\rm e}^{-i\hat{\rm H}T/\hbar}|{\bf N}_{i} \rangle =
{\rm e}^{-iS_{0}(\phi_{f}-\phi_{i})} \int {\cal D}[\theta]{\cal D}
[\phi] \exp \left( \frac{i}{\hbar}\int^{t_{f}}_{t_{i}}{\cal L}^{\prime}  
{\rm d}t \right) ,\\
{\cal L}^{\prime}=\Omega \left[ \frac{m}{\gamma} \dot \phi \cos \theta 
+\frac{\chi_{\bot}}{2\gamma^{2}}(\dot{\theta}^{2} +\dot{\phi}^{2} \sin^{2}
\theta) \right]-V(\theta, \phi).
\end{eqnarray}
Where $(\theta_{1},\phi_{1})$ has been replaced by $(\theta, \phi)$ .
$V(\theta,\phi)=\Omega (K_{\bot} \cos^{2} 
\theta+K_{\|} \sin^{2} \theta \sin^{2}\phi-mH \sin \theta 
\sin \phi)$, $S_{0}=S_{1}+S_{2}\;,\;  m=\gamma\hbar (S_{1}-S_{2})/\Omega$,   
$\Omega$ is the volume of the AFM particle. $K_{\bot}=2k_{\bot}S^{2}/\Omega$  
and  $K_{\|}=2k_{\|}S^{2}/\Omega$ are the trasnverse and longitudinal anisotropy 
constants, respectively. We have set $S_{1}=S_{2}=S$ except in the terms 
containing $S_{1}-S_{2}$.  The parameter $\chi_{\bot}=\gamma^{2}/J$ is
introduced according to Ref.[20] for the problem at hand. 
 
We consider a very strong transverse anisotropy $i.e. \;\; K_{\bot} \gg K_{\|}$. 
In this case  N$\acute{\rm e}$el vector is forced to lie in the x-y plane. Replacing  $ \theta$
by $\pi/2+\eta_{\theta}$ where $\eta_{\theta}$ denotes the small fluctuation and 
carrying out integral over $\eta_{\theta}$ we obtain
\begin{equation}
\langle {\bf N}_{f}|{\rm e}^{-i\hat{\rm H}T/\hbar}|{\bf N}_{i} \rangle={\rm e}^{-iS_{0}
(\phi_{f}-\phi_{i})} \int {\cal D}[\phi]\exp \left( \frac{i}{\hbar} \int
^{t_{f}}_{t_{i}}{\cal L}_{eff}{\rm d}t \right), 
\end{equation}
where
\begin{equation}
{\cal L}_{eff}=(I_{f}+I_{a})\frac{\Omega}{2} \left( \frac{{\rm d}\phi}
{{\rm d}t}\right) ^{2}- V(\phi)
\end{equation}
is the effective  Lagrangian which is seen to be the Lagrangian of a plane rotor. Where $I_{f}=m^{2}/2\gamma^{2}
K_{\bot}$ and $I_{a}=\chi_{\bot}/\gamma^{2}$ are the effective FM and AFM 
moments of  inertia, respectively[6]. $V(\phi)=\Omega K_{\|}(\sin \phi-
\triangle)^{2} \;( \triangle =H/H_{c}$ with a parameter 
$H_{c}=2K_{\|}/m$) is the effective potential. It is seen that the  
net magnetic moment of the noncompensated sublattices in the applied 
magnetic field shifts the equilibrium orientations of  N$\acute{\rm e}$el 
vector for corresponding angles $\pm \arcsin\triangle$ as shown in Fig.1-(b)
besides the modification of FM moment of inertia $I_{f}$ given in Refs.[6,18].
It may be worth while to compare
our results with that in literature. In the  absence of the magnetic field 
(namely $\triangle=0$) the two degenerate equilibrium orientations return
to the positive and negative ${\rm x}$-axis (see Fig.1-(a)) respectively 
in agreement with the equilibrium phases of the AFM particle  with noncompensated 
sublattices[18]. The small oscillation frequency of  N$\acute{\rm e}$el vector 
around its equilibrium orientations which serves as a characteristic parameter 
for the flip of  N$\acute{\rm e}$el vector of the AFM particle is seen to be 
\begin{equation}
\omega(H,m)=\left [ \frac{2K_{\|}(1-\triangle^{2})}{I_{f}+I_{a}}\right ]
^{1/2}.
\end{equation}
For $\triangle=0$ it reduces exactly to 
\begin{equation}
\omega(H=0,m)=\left [ \frac{2K_{\|}}{I_{f}+I_{a}}\right ]^{1/2}
\end{equation}
as given in Ref.[18]. If we consider idealized sublattices that $m=0$ the frequency 
Eq.(17) goes back to the well known value
\begin{equation}
\omega(H=m=0)=\gamma\left [ \frac{2K_{\|}}{\chi_{\bot}}\right ]^{1/2}=
\left [ 2 K_{\|}J \right ]^{1/2}.
\end{equation}
The effective potential of the plane rotor is plotted in Fig.2. The minima of 
the potential correspond to the equilibrium orientations of the  
N$\acute{\rm e}$el vector. The energy of net magnetic moment in the applied    
magnetic field lowers the barrier height in the direction of magnetic field
while increases the barrier height in the opposite direction. We are interested in the quantum tunneling of the 
effective plane rotor through the barriers.

\section
{ Two types of instantons and level shifts}

In order to obtain the tunneling rate we evaluate the Euclidean path integrls in Eq.(15)
with  the Wick rotation $t=i\tau$ .
The Euclidean Lagrangian for the pseudoparticle moving in the classical 
forbidden region, namely, in the barrier is seen to  be
$L_{E}=(I_{f}+I_{a})\frac{\Omega}{2}\left(\frac{{\rm d}\phi}{{\rm d}\tau}\right)^{2}
+V(\phi)$. The equation of motion  of rotor at finite energy is 
 \begin{equation}
\frac{\Omega}{2} ( I_{f}+I_{a} ) \left( \frac{{ \rm d} \phi}{{ \rm d} \tau}  
 \right)^{2}-V(\phi) =-E\;,
\end{equation}
The periodic potential $V(\phi)=V(\phi+2n\pi)$ has an asymmetric  twin barrier
(Fig.2) .The two-fold rotation symmetry[10], namely, $V(\phi+\pi)=V(\phi)$ 
in the presence of magnetic field is removed . When the energy is higher than 
the ground sta8te tunneling is dominated by periodic instantons[21,22].
Thus there are two different periodic instantons
corresponding to two types of barriers. With the periodic boundary condition, two periodic instanton 
solutions of Eq.(17) are found to be
\begin {equation}
 \phi^{(1)}_{c}=\frac{\pi}{2}+2\arctan [\lambda_{1} {\rm sn} (q\tau, k) ]\;,
 \end{equation}
 \begin{equation}
 \phi^{(2)}_{c}=\frac{3\pi}{2}-2\arctan [\lambda_{2} {\rm sn} (q\tau, k) ]\;,
\end{equation}
where ${\rm sn}(q\tau, k)$ is elliptic function with modulus $k$ and  period
$4\kappa (k)$. $\kappa  (k)$  denotes the complete elliptic integral of the first
kind, with
$$ 
 k=\left [ \frac{(1-\eta)^{2}-\triangle^{2}}
 {(1+\eta)^{2}-\triangle^{2} } \right]^{1/2}\;,\;\;\;
 \eta=\left ( \frac{E}{\Omega K_{\|} } \right )^{1/2} \;.
 $$
The parameters $q\,, \lambda_{1}$ and $\lambda_{2}$ are defined by
 $$
 q=\left( \frac{K_{\|}}{2(I_{f}+I_{a})} \right )^{1/2} [ 
 (1+\eta) ^{2} -\triangle^{2} ]^{1/2}\;,\;\;\;
 \lambda_{i}=\left [ \frac{(1-\eta)^{2}-\triangle^{2}}{(1-(-1)^{i}
 \triangle)^{2}-\eta^{2}} \right ]^{1/2}\;,\;\;(i=1,2)
$$
The trajectories  of instantons $ \phi^{(1)}_{c}$ and $\phi^{(2)}_{c}$ are 
shown in Fig.2. At initial time $\tau_{i}$, instantons $\phi^{(1)}_{c}$, 
 $\phi^{(2)}_{c}$ start from the potential well at $ \phi_{i}=\arcsin \triangle $ 
and reach the neighbouring well at $\phi_{f}=\pi-\arcsin \triangle$
at final time $\tau_{f}$ along the anticlockwise (through small barrier) and   
clockwise (through large barrier) paths respectively. In other words
 the N$\acute{\rm e}$el vector tunnels
through a large  barrier (or small barrier) between two angular positions with  
the lowest energy.

We assume that $|m,\phi^{(1)}_{2n} \rangle $ and  $|m,\phi^{(2)}_{2n+1} \rangle$ denote the 
eigenstates of  the harmonic oscillator approximated Hamiltonian in the 
potential wells at $\phi^{(1)}_{2n}=2n\pi+\arcsin \triangle $ and 
$\phi^{(2)}_{2n+1}=(2n+1)\pi-\arcsin \triangle $, respectively, where $m$ is the  
index of low-lying levels.  The amplitudes tunneling through two different 
barriers are given by[21] 
\begin{equation}
A^{(1)}_{m}=\langle m, \phi^{(1)}_{0}|{\rm e}^{-\frac{2\beta \hat{\rm H}}{\hbar}}
|m, \phi^{(2)}_{1}\rangle =\exp \left(-\frac{2\beta \varepsilon_{m}}{\hbar}
\right) \sinh\left(  \frac{2\beta \triangle \varepsilon^{(1)}_{m}}{\hbar} 
\right)  \;,
\end{equation}
\begin{equation}
A^{(2)}_{m}=\langle m, \phi^{(2)}_{1}|{\rm e}^{-\frac{2\beta \hat{\rm H}}{\hbar}}
|m, \phi^{(1)}_{2}\rangle =\exp \left(-\frac{2\beta \varepsilon_{m}}{\hbar}
\right)\sinh\left(  \frac{2\beta \triangle \varepsilon^{(2)}_{m}}{\hbar} 
\right) \;,
\end{equation}
where $\triangle \varepsilon^{(1)}_{m} (\triangle \varepsilon^{(2)}_{m}) $
is the level shift induced by tunneling through the small (large) 
barrier alone. $\triangle \varepsilon^{(i)}_{m}$ is actually the overlap  
integral defined in the following Eqs.(42),(43). Where $\beta=\tau_{f}-
\tau_{i}$. With  the help of the path integral, the matrix element in Eq.(23) and Eq.(24) 
can be rewritten as 
\begin{equation}
A^{(1)}_{m}=\int \psi^{*}_{m}(\phi^{(1)}_{0}, \phi_{f})\psi_{m}(\phi^{(2)}_{1},
\phi_{i}){\cal K}^{(1)}(\phi_{f}, \tau_{f};\;\phi_{i},\tau_{i}){\rm d}\phi_
{f}{\rm d}\phi_{i}\;,
\end{equation} 
\begin{equation}
A^{(2)}_{m}=\int \psi^{*}_{m}(\phi^{(2)}_{1}, \phi_{f})\psi_{m}(\phi^{(1)}_{2},
\phi_{i}){\cal K}^{(2)}(\phi_{f}, \tau_{f};\;\phi_{i},\tau_{i}){\rm d}\phi_
{f}{\rm d}\phi_{i}\;,
\end{equation} 
where 
$$ \psi^{*}_{m}(\phi^{(1)}_{0}, \phi_{f})=\langle m, \phi^{(1)}_{0}|
   \phi_{f} \rangle\;,\;\;
   \psi_{m}(\phi^{(2)}_{1}, \phi_{i})=\langle \phi_{i} | 
   m, \phi^{(2)}_{1} \rangle\;,  $$
$$ \psi^{*}_{m}(\phi^{(2)}_{1}, \phi_{f})=\langle m, \phi^{(2)}_{1}|
   \phi_{f} \rangle\;,\;\;
   \psi_{m}(\phi^{(1)}_{2}, \phi_{i})=\langle \phi_{i}   |
   m, \phi^{(1)}_{2} \rangle\;,   $$
\begin{equation}
{\cal K}^{(i)}(\phi_{f}, \tau_{f};\;\phi_{i},\tau_{i}) =\int ^{\phi_{f}}_
{\phi_{i}}{\cal D}[\phi]\exp \left( -\frac{S^{(i)}_{E}}{\hbar}  \right)\;,
\end{equation}
\begin{equation}
S^{(i)}_{E}=\int ^{\tau_{i}}_{\tau_{f}} \left[ \frac{\Omega}{2} ( I_{f}+I_{a}) 
\left( \frac{{ \rm d} \phi}{{ \rm d} \tau} \right)^{2} +V(\phi) \right]  {\rm d}
\tau \;,\;\; (i=1,2).
\end{equation}
Here $S^{(i)}_{E}$ denotes the Euclidean action, and ${{\cal K}^{(i)}(\phi_{f}, 
\tau_{f};\;\phi_{i},\tau_{i}) }$ is Feynman propagator through two kinds of 
barriers. Substituting the periodic instanton solutions Eq.(21) and Eq.(22) into 
Eq.(28) the Euclidean action along the classical trajectory is found to be 
\begin{equation}
S^{(i)}_{Ec}=2E\beta+W^{(i)}\;,
\end{equation}
\begin{equation}
W^{(i)}=\frac{4\Omega \chi_{\bot}q}{\gamma^{2}\lambda^{2}_{i}}\left[\lambda
^{2}_{i}E(k)+(k^{2}-\lambda^{2}_{i})\kappa(k)+(\lambda^{4}_{i}-k^{2})\Pi(k,
\lambda_{i})\right]\;,
\end{equation}
where $E(k)$ is the complete elliptic integral of the second kind, and  $\Pi
(k,\lambda_{i})$ is the complete elliptic integral of the third 
kind with the parameter $\lambda_{i}$. The level shift $\triangle \varepsilon ^
 {(i)}_{m}$ can be determined by completing the integrals of (25) and (26) and 
  comparing the result with Eq.(23) and Eq.(24). Using the method  in Ref.[21], 
  the transition amplitude is obtained as
  \begin{equation}
  A^{(i)}_{m}=\exp \left( -\frac{2E\beta}{\hbar} \right) \sinh \left \{ 
  \frac{\beta}{\sigma \kappa(k')} \left( \frac{2K_{\|}}{I_{a}+I_{f}} \right)^
  {1/2} \exp \left(-\frac{W^{(i)}}{\hbar} \right)  \right\} \;,\;\;(i=1,2),
  \end{equation}
where $k'=(1-k^{2})^{1/2}$ and $\sigma=[(1+\eta)^{2}-\triangle^{2}]^
{1/2}$. Comparing this expression with Eq.(23) and Eq.(24) we find that 
two types of level shifts are 
\begin{equation}
  \triangle \varepsilon^{(i)}_{m}=\frac{\hbar}{\sigma \kappa(k')} \left( 
  \frac{2K_{\|}}{I_{a}+I_{f}}  \right)^{1/2} \exp \left(-\frac{W^{(i)}}{\hbar} 
  \right),\;\;(i=1,2),
\end{equation}

\section
{The magnetic field dependence of tunneling rates }

It is easy to see that the difference between the heights of larger and small 
barriers increases with the external magnetic field .  On the other hand, the 
effective frequency of oscillator near the bottom of potential well given in Eq.(17)
$i.\,e.$ $\omega=\omega_{0}(1-\triangle^{2})^{1/2}$ decreases with the
increasing field, where $\omega_{0}=(\frac{2K_{\|}}{I_{a}+I_{f}})^{1/2}$ is the 
frequency in the absence of the magnetic field[18]. For the 
low energy case that the energy $E$ is far below the barrier height , 
 $i.e.\; \eta \ll 1-\triangle,
   k\rightarrow 1,\; k'\rightarrow 0$, $E(k)$, $\kappa(k)$ can be expanded
   as power series of $ k^{\prime}$. The complete elliptic integral of the third kind is 
   expressed as 
   $$
   \Pi(k,\lambda_{i})=\frac{k^{2}}{k^{2}+\lambda_{i}^{2}}+\left[ \frac{\lambda
   ^{2}_{i}}{(1+\lambda^{2}_{i})(k^{2}+\lambda^{2}_{i})} \right]^{1/2} \left\{
   E(k)F(\alpha^{i},k^{\prime})+\kappa(k)[E(\alpha^{i},k')-F(\alpha^{i},k')] \right\},
   $$
where $\alpha^{(i)}=\arcsin (\lambda^{2}_{i}/(k^{2}+\lambda^{2}_{i}))^{1/2}.\;\;
F(\alpha^{(i)},k'),\;\;E(\alpha^{(i)},k')$ are the incomplete elliptic integrals 
of the first and second kinds, respectively. Thus $\Pi(k, \lambda_{i})$ can be 
also expanded as the power series of $k'$. Then, the power series expression 
of $W^{(i)}$ reads 
\begin{equation}
W^{(1)}=\frac{4K_{\|}\Omega}{\omega_{0}} \left\{(1-\triangle^{2})^{1/2}-
\triangle \arccos \triangle -\frac{1}{16}(1-\triangle^{2})^{3/2}k'^{4} \left[ 
\ln \frac{4}{k'}+1 \right] \right\} \;,
\end{equation}
\begin{equation}
W^{(2)}=\frac{4K_{\|}\Omega}{\omega_{0}} \left\{(1-\triangle^{2})^{1/2}+
\triangle \arccos (-\triangle) -\frac{1}{16}(1-\triangle^{2})^{3/2}k'^{4} \left[ 
\ln \frac{4}{k'}+1 \right] \right\} \;,
\end{equation}
In the low energy case, $k'=4\eta/(1-\triangle^{2})\ll1$,we may take the  oscillator 
approximated energy-quantization $i.e.\; E\rightarrow E_{m}=(m+1/2)\hbar 
\omega$. Taking note of limits $\kappa(k\rightarrow 0)\rightarrow \pi/2 \; $  and  $\;\sigma(\eta
 \rightarrow 0) \rightarrow (1-\triangle^{2})^{-1/2} $ we find
 \begin {equation}
 \triangle\varepsilon^{(1)}_{m}=F_{m}(1-\triangle^{2})^{5/4+3m/2}\exp \{-B[(1-
 \triangle^{2})^{1/2}-\triangle \arccos \triangle] \},
\end{equation}
 \begin {equation}
 \triangle\varepsilon^{(2)}_{m}=F_{m}(1-\triangle^{2})^{5/4+3m/2}\exp \{-B[(1-
 \triangle^{2})^{1/2}+\triangle \arccos (-\triangle)] \}.
 \end{equation}
Where
$$  F_{m}=F_{0}\frac{1}{n!}[4B]^{n}
 \;,\;B=\frac{4K_{\|} \Omega}{\omega_{0}\hbar} \;, F_{0}=\hbar\omega_{0}\left[
\frac{8B}{\pi}\right]^{1/2}. $$
Eq.(35) and Eq.(36) give rise to the field dependence of the level shift for 
low-lying excited states. There is an obvious difference  between the  
 level shifts induced by tunneling through two kinds of barriers. For a given 
 excited state, $\triangle \varepsilon^{(i)}_{m}$ as a function of the external 
 magnetic field is plotted in Fig. 3, with $k_{\|}=10^{6} erg/cm^{3}\;,\;
 K_{\bot}=10^{8} erg/cm^{3}\;,\; \chi_{\bot}=10^{4}\;,$ the excess of spin $S_{1}-S_{2}=10$
 and the AFM particle radius $r=7.5 nm$. It is clearly shown that the 
 tunneling rate through a small barrier increases rapidly with the 
 external magnetic  field because the field reduces both the height and width
 of barrier. The situation is just  opposite for the tunneling through the 
larger barrier . In addition, Fig.3 also shows that the tunneling rate increases 
with the energy  levels in the low-lying excited  states. When $\triangle=
H/H_{c}$ attends to 1 there is no tunneling at all since the 
small barrier shrinks to zero and only one easy direction remains. In the 
absence of applied magnetic field we have $(\triangle=0) \triangle \varepsilon^{(1)}_{m}=\triangle \varepsilon^{(2)}_{m}$
and for the ground state tunneling,namely $E=\eta=0$,the level shift $\triangle \varepsilon_{0}$
reduces to exactly the result in Ref.[18].

\section
{Level splitting and quantum interference effect}

$\triangle \varepsilon^{(i)}_{m}$ is only the level shift induced by tunneling
through a single barrier (smaller or larger). The  periodic potential 
 $V(\phi)=V(\phi+2n\pi)$ can be regarded as a one-dimensional superlattice
consisting of two sublattices. The general translation  symmetry results in 
the energy band structure, and the energy spectrum could be determined with 
the  Bloch theory.  
Let $|m,\phi^{(1)}_{2n} \rangle$ be the eigenstates of the zero order Hamiltonian  
$\hat{\rm H}^{(1)}_{0}$ in  the potential well which lies at $\phi^{(1)}_{2n}=
2n\pi+\arcsin \triangle$. $|m, \phi^{(2)}_{2n+1} \rangle$ denote the eigenstates 
of the zero order Hamiltonian ${\rm H}^{(2)}_{0}$ in the well at 
$\phi^{(2)}_{2n+1}=(2n+1)\pi-\arcsin \triangle $. Thus  
\begin{eqnarray}
{\rm H}^{(1)}_{0} |m, \phi^{(1)}_{2n} \rangle   & = & \varepsilon_{m} |m, \phi^{(1)}_
						 {2n} \rangle\;,  \\
{\rm H}^{(2)}_{0} |m, \phi^{(2)}_{2n+1} \rangle & = & \varepsilon_{m} |m, \phi^{(2)}_
						   {2n+1} \rangle\;.
\end{eqnarray}
Bloch state with $2\pi$ periodic boundary condition is written as 
\begin{equation}
 |\psi\rangle=\sum_{n} \left( {\rm e}^{i(\xi+S_{0})\phi^{(1)}_{2n}}
|m, \phi^{(1)}_{2n} \rangle +  {\rm e}^{i(\xi+S_{0})\phi^{(2)}_{2n+1}}
|m, \phi^{(2)}_{2n+1} \rangle \right)\;,
\end {equation}
where $\xi$ is Bloch wave vector , $ {\rm e}^{iS_{0}\phi^{(1)}_{2n}}$(or 
${\rm e}^{iS_{0}\phi^{(2)}_{2n+1}}$) is seen to be the topological phase from
Eq.(15). Substituting  Eq.(39) into the following stationary 
Schr$\ddot{o}$dinger equation 
\begin{equation}
\hat{\rm H}|\psi\rangle=E|\psi\rangle\;,
\end{equation}
and taking into account only the nearest neighbours  yield the energy spectrum as
\begin{eqnarray}
E&=& \varepsilon_{m}- \triangle \varepsilon^{(1)}_{m}\cos [(\xi+S_{0})(\pi-2\arcsin
    \triangle)]  \nonumber  \\
 & &  -\triangle \varepsilon _{m}^{(2)}\cos [(\xi+S_{0})(\pi+2\arcsin\triangle)]  ,
\end{eqnarray}
where the level shift $\triangle \varepsilon^{(i)}_{m}$ is actually the overlap 
integral defined by
\begin{eqnarray}
\triangle \varepsilon^{(1)}_{m} & = &- \int u^{*}_{m}(\phi-\phi^{(1)}_{2n}){\rm H}u_{m}
				      (\phi-\phi^{(2)}_{2n+1}) {\rm d}\phi\;,\\
\triangle \varepsilon^{(2)}_{m} & = & -\int u^{*}_{m}(\phi-\phi^{(1)}_{2n}){\rm H}u_{m}
				      (\phi-\phi^{(2)}_{2n-1}){\rm d}\phi\;.
\end{eqnarray}
The Bloch wave vector $\xi$ can be assumed to take either of the two values 0 and 1 in the first Brillouin 
zone[21,23]. Thus the level splitting is seen to be 
\begin{eqnarray}
 \triangle 
\varepsilon_{m} & = & |\triangle\varepsilon_{m}^{(1)}\cos [2(1+S_{0})
                         \arccos \triangle]+(-1)^{2S_{0}} \triangle\varepsilon_{m}^{(2)}
			  \cos [2(1+S_{0})\arccos\triangle]    \nonumber  \\
	       &   &    -\triangle\varepsilon_{m}^{(1)}\cos [2S_{0}
                         \arccos \triangle]- (-1)^{2S_{0}}\triangle\varepsilon_{m}^{(2)}
			  \cos [2S_{0}\arccos\triangle]|    \nonumber  \\
  & = & E^{\pm}_{m}|\sin [(2S_{0}+1)\arccos \triangle]|\;,
\end{eqnarray}
where
\begin{eqnarray}
E^{+}_{m} & = & R_{m} \cosh(B\pi \triangle/2)\;,\;\;(S_{0}={\rm integral})\;,     \\
E^{-}_{m} & = & R_{m} \sinh(B\pi \triangle/2)\;,\;\;(S_{0}={\rm half-integer})\;,\\
    R_{m} & = & 4F_{m}(1-\triangle^{2})^{7/4+3m/2} \nonumber  \\
	  &   &  \exp\{-B[(1-\triangle^{2})
	       ^{1/2}+\triangle\arcsin \triangle]\}.
\end{eqnarray}

When $H=0, i.e.\arcsin\triangle=0$, energy 
spectrum in Eq.(41) reduces to  the result in Ref.[14], and for $S_{0}=$ half-integer 
the MQC is quenched in agreement with Kramers' theorem which can be well understood 
as two-fold discrete rotation-symmetry of Hamiltonian[10] . The applied magnetic 
field breaks the rotation symmetry and the Kramers' degeneracy is 
removed. Level splitting Eq.(44) shows the quantum interference  effect  depending 
on  applied magnetic field. The level splitting 
increases with the magnetic field and whenever  the  magnetic field reaches  some 
specific values which satisfy $H/H_{c}=\cos [l\pi /(2S_{0}+1)]$ ($l$ is a 
integer), we have $ \triangle \varepsilon_{m}=0$ no matter $S_{0}$ is integer or  
half-integer. The quenching is similar to 
the case of the FM  particle in Refs.[9,24] and is 
the result of quantum interference between transitions through  the two kinds 
of barriers.  Fig.4 shows the oscillation of the level 
splitting with the field for the ground state.  

\section
{Conclusion}

We present a full study of quantum tunneling effect for AFM particle with a 
small  non-compensation of sublattices in an external magnetic field. The level
splitting which is obtained only for ground state in literature has been 
extended to low-lying excited states with the help of periodic instanton method. 
The quantum interference effect, particularly, the entire suppression of tunneling 
may be of significance for practical application.
\vspace{1cm}
\begin{center}
Acknowledgments
\end{center}
This work was supported by the 
National Natural Science Foundation of China under Grant No. 19775033.

\newpage
\begin{flushleft}          
{\Large  \bf References}
\end{flushleft}
\begin{flushleft}
\begin{enumerate} {\small \def\baslinestretch{1.3}
\item M. Enz and R. Schilling, J. Phys.C: Solid State Phys. 19(1986)L711.
\item E. M. Chudnovsky and L. Gunther, Phys. Rev. Lett. 60(1988)661.
\item E. M. Chudnovsky and L. Gunther, Phys. Rev. B 37(1988)9455.
\item P. C. E. Stamp, Phys. Rev. Lett. 66(1991)2802.
\item B. Barbara and E. M. Chudnovsky, Phys. Lett. A 145(1990)205.
\item E. M. Chudnovsky, J. Magn. Magn. Matter, 140-144(1995)1821.
\item J. M. Duan and A. Garg, J. Phys.: Condense Matter, 7(1995)2171.
\item D. Loss, D. P. Divincenzo and G. Grinstein, Phys. Rev. Lett. 69(1992)3232.
\item A. Garg. Europhys. Lett. 22(1993)205.
\item B. Z. Li,  J. H. Wu, W. D. Zhong and F. C. Pu, Science in China, (series A),
      2(1998)321.
\item Rong L$\ddot{{\rm u}}\;$, J.-L. Zhu and L. Chang,  Phys. Rev. B 56(1997)10933.
\item X. B. Wang and F. C. Pu, J. Phys.: Condense Matter. 8(1996)L541. 
\item I. V. Krive and O. B. Zaslavkii, Phys.: Condense Matter, 2(1990)9457.
\item J.-Q. Liang and H. J. W. M$\ddot{{\rm u}}$ller-Kirsten, Jian-Ge Zhou, Z. Phys. B 
      102(1997)525.
\item H. Simanjuntak, J. Phys.: Condense matter, 6(1994)2925.
\item J.-Q. Liang, H. J. W. M$\ddot{{\rm u}}$ller-Kirsten,Y.-H. Nie, F.-C. Pu, Y.-B. Zhang,
      Phys. Lett. A248(1998)434.
\item J. Tejada, X. X. Zhang, J. Phys.: Condense Matter, 6(1994)263.
\item E. M. Chudnovsky, "Quantum Tunneling of Magnetization--QTM'94",  pp. 77-89, 
      Eds. L. Gunther and B. Barbara.
\item E. Frandkin, Field Theories of Condensed Matter Systems (Addison-Wesley,
       Redwood City) 1991, Chapt.5.
\item J. V. Delft and G. L. Henley, Phys. Rev. Lett. 69(1992)3236.
\item J.-Q. Liang, H. J. W. M$\ddot{{\rm u}}$ller-Kirsten, Jian-Ge Zhou , F. Zimmerschied, 
       and F. C. Pu, Phys. Lett. B 393(1997)368.
\item J.-Q. Liang and H. J. W. M$\ddot{{\rm u}}$ller-Kirsten, Phys. Rev. D 46(1992)
      4685, D 50(1994)6519, D51(1995)718.
\item J.-Q. Liang,Y.-B.Zhang, H. J. W. M$\ddot{{\rm u}}$ller-Kirsten, Jian-Ge Zhou ,
      F. Zimmerschied, and F.-C. Pu, Phys. Rev. B 57(1998)529.

\item E. M. Chudnovsky and David P. DiVincenzo,  Phys. Rev. B 48(1993)10548.}

\end{enumerate}
\end{flushleft}

\newpage
\begin{center}
Figure captions
\end{center}

Fig.1 (a) The two degenerate equilibrium orientations of  N$\acute{\rm e}$el 
          vector with noncompensated sublattices in an applied magnetic field.   
          $\phi_{+}=\arcsin\triangle$.
      (b) The equilibrium orientatons of N$\acute{\rm e}$el vector in the absence 
          of applied magnetic field ($\triangle=0$).
	  
Fig.2  The periodic potential with asymmetric twin-barrier and the instanton 
       trajectories.

Fig.3  The level shift $\triangle\varepsilon^{(i)}_{m}$ as a function of $H/H
      _{c}$. Solid line for $\triangle\varepsilon^{(1)}_{m}$ and dotted line for $\triangle\varepsilon^{(2)}_{m}$.       

Fig.4  The level splitting at ground state as a function of $H/H_{c}$.

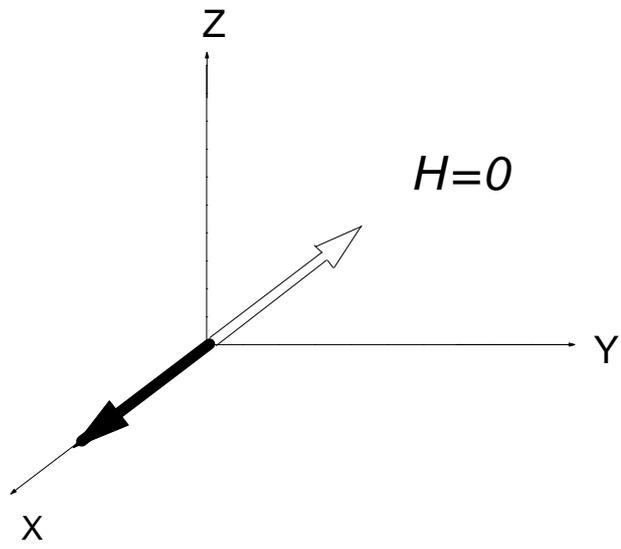
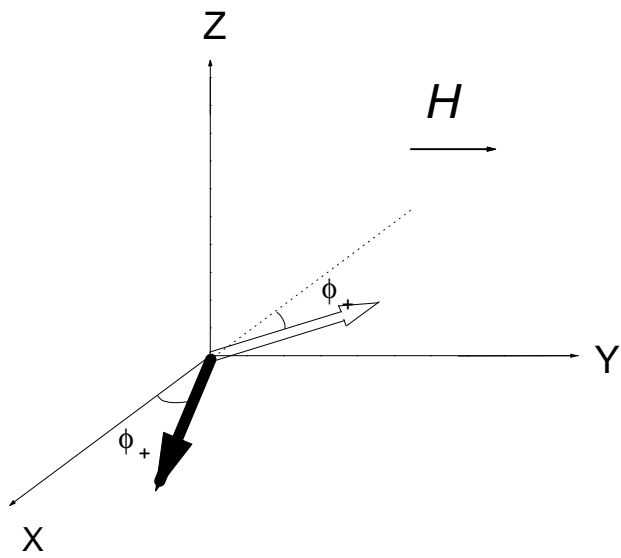
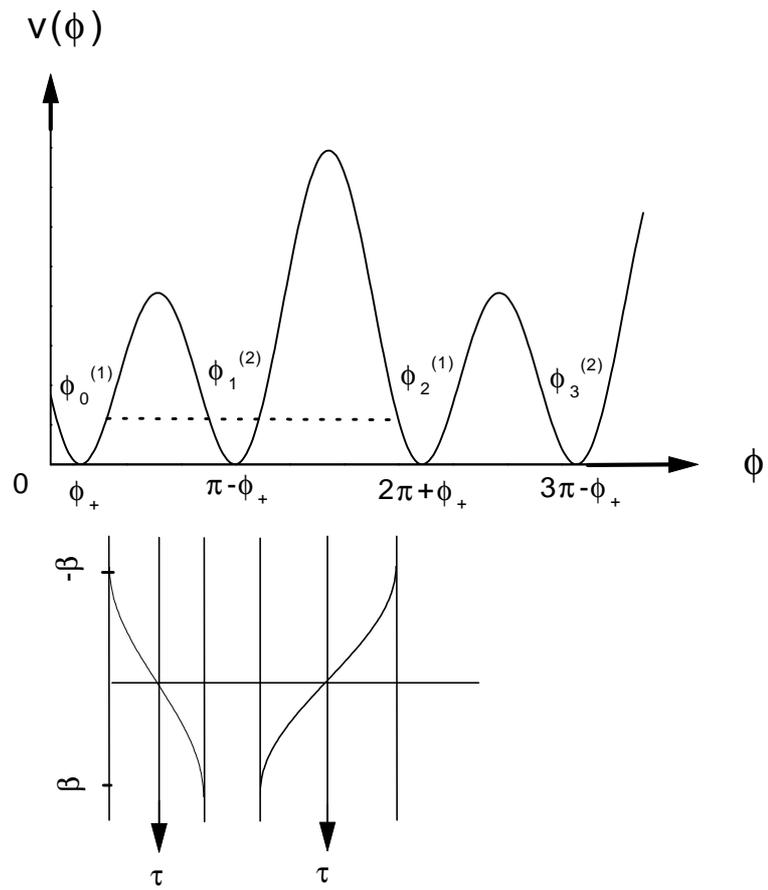
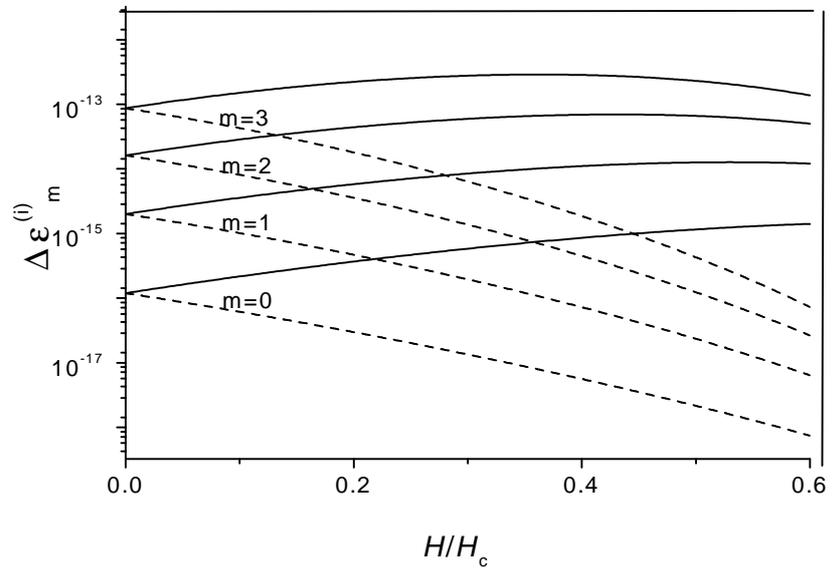
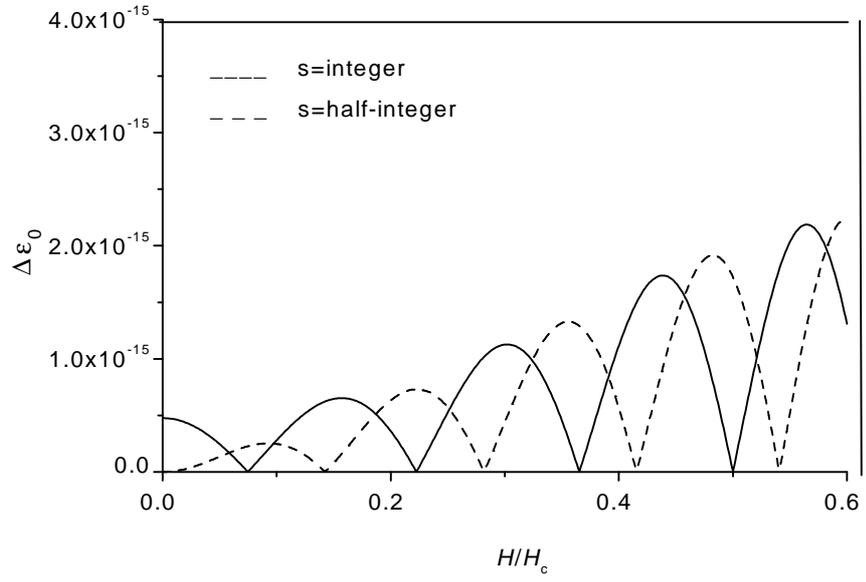
\end{document}